# Anomalous behavior of acoustic phonon mode and central peak in Pb(Zn$_{1/3}$Nb$_{2/3}$)$_{0.85}$Ti$_{0.15}$O$_3$ single crystal studied using Brillouin scattering

K. K. Mishra[1*], V. Sivasubramanian[1], A.K. Arora[1] and Dillip Pradhan[2]

[1]Condensed Matter Physics Division, Indira Gandhi Centre for Atomic Research, Kalpakkam, 603102, India

[2]Department of Physics, NIT Rourkela, 769008, India

## Abstract

Brillouin spectroscopic measurements have been carried out on relaxor ferroelectric Pb(Zn$_{1/3}$Nb$_{2/3}$)$_{0.85}$Ti$_{0.15}$O$_3$ (PZN-PT) single crystal over the temperature range 300-585 K. The longitudinal acoustic phonon begins to soften below 650 K, which is attributed to the Burns temperature ($T_B$). On the other hand, the line width of the longitudinal acoustic (LA) phonon mode exhibits a sharp Landau-Khalatnikov-like maximum and an accompanying anomaly in the LA mode frequency around 463 K, the tetragonal-cubic phase transition temperature ($T_{tc}$). In addition, a broad central peak, found below the characteristic intermediate temperature $T^* \sim 525$ K, exhibits critical slowing down upon approaching $T_{tc}$ indicating an order-disorder nature of the phase transition. The relaxation time of polar nano regions estimated from the broad central peak is found to be same as that obtained for LA phonon mode suggesting an electrostrictive coupling between strain and polarization fluctuations. The activation energy for the PNRs relaxation-dynamics is found to be ~236 meV. Polarized nature of the central peak suggests that the polar nano regions have the tendency to form long-range polar ordering.

# I. INTRODUCTION

Lead-based relaxor ferroelectric materials have attracted the scientific community from the point of view of fundamental physics as well as industrial application.[1-6] Relaxor ferroelectrics are characterized by high dielectric constant, slim hysteresis loop, and broad frequency and temperature dependent dielectric maximum indicative of multiple scales of relaxation. In order to understand the microscopic origin of the complex behavior of relaxor ferroelectric materials different theoretical models have been proposed.[4-6] It is widely accepted that the central feature to the complex behavior in these materials is the nucleation and growth of the polar nano regions (PNRs). The PNRs possess local and randomly oriented finite ferroelectric polarizations. Obtaining an understanding of such nanoscopic PNRs and their consequences on macroscopic properties such as ferroelectric phase transition has been addressed in many studies.[7-12] While cooling from the paraelectric phase at high temperature, PNRs begin to nucleate below a temperature, known as Burns temperature $T_B$.[13] Because of formation of PNRs a deviation from Curie-Weiss law is also found below $T_B$.[14-16] In order to understand the dynamical aspects of PNRs many investigations such as dielectric constant, neutron scattering, far-infrared transmission spectroscopy, Raman and Brillouin spectroscopy have been carried out on many relaxor systems.[7,8,17-22] A central peak in the Brillouin spectrum has been reported to arise due to relaxation dynamics of PNRs. In addition, other manifestation of dynamical features is the coupling of strain and polarization fluctuation leading to a softening of the acoustic phonon.[17-20]

In the present work, we focus on the Brillouin spectroscopic study on tetragonal-cubic phase transition and the dynamical behaviour of the PZN-PT single crystal over a wide range of temperature. The phase diagram of the (1-$x$)PZN-($x$)PT solid solution shows a sequence of structural phases with temperature and PT content.[23] Brillouin studies on (1-$x$)PZN-($x$)PT below and around the morphotropic phase boundary (MPB) composition[23] have been reported earlier.[7, 8, 24, 25] $x$ = 4.5[9] and 7%[8] Ti-doped system show typical relaxor character, whereas occurrence of long range ferroelectric order was found at $x$ = 9%.[9] However, the solid solution far away from MPB such as that $x$ = 15% has not been investigated yet. Since the increase in Pb content influences the structural[23] and relaxor behaviour of PZN-PT,[26, 27] it is also expected to modify the central peak and the acoustic phonon mode. Furthermore, it is of interest to examine the differences in the behaviours of solid solutions near and away from MPB. Such study on a broad compositional range is expected to improve the understanding of relaxation dynamics of PNRs. The aim of the present work is to study the anomalous behaviour of acoustic phonon and central peak across tetragonal-cubic phase transition and the dynamical behaviour of PNRs in Pb(Zn$_{1/3}$Nb$_{2/3}$)$_{0.85}$Ti$_{0.15}$O$_3$ (PZN-PT) single crystal. Single crystals were synthesized by flux method and relaxor-like behavior was examined using dielectric spectroscopy.

## II. EXPERIMENTAL

Transparent and yellow coloured single crystals of Pb(Zn$_{1/3}$Nb$_{2/3}$)$_{0.85}$Ti$_{0.15}$O$_3$ were grown by flux method with Pb$_3$O$_4$ as flux. The charge and flux were taken in 40:60 ratio by weight. The crystals were grown by cooling from 1473 to 1203 K at a rate of 0.8 K / h and then from 1203 K to room temperature at a rate of 300 K / h. The Perovskite phase was confirmed using X-ray diffraction analysis. The typical sample dimensions were 3 × 2 ×0.5 mm$^3$. From the Laue pattern the single crystals were found to be oriented in [100] plane. The surfaces were polished to optical quality. The Energy dispersive X-ray analysis of the sample was carried out using a scanning electron microscope (CARL ZEISS, SUPRA 55). The dielectric parameters (capacitance and dissipation factor) were measured in a wide range of frequency (100 Hz to 1 MHz) using a computer–controlled

LCR meter/ impedance analyzer (PSM: N4L (Model: 1735, UK)) in the temperature range 300-585 K. Brillouin spectra were measured in backscattering geometry using a high-contrast 3+3-pass tandem Fabry-Perot interferometer (JRS Scientific Instruments). The sample was excited using 532 nm light using a single-mode diode-pumped solid state laser (Diabolo 500). The spectra were recorded using the free spectral ranges (FSR) of 100 and 400 GHz in the temperature range 300 - 873 K using a commercially available heating stage with a stability of ± 0.1 K. The exact sample temperature was monitored using a thermocouple placed adjacent to the sample inside the heating stage. The measurements were performed at successively higher *T* during a heating cycle.

## III. RESULTS AND DISCUSSION

Figure 1 shows the X-ray diffraction pattern of the powdered crystals of PZN-PT. All the diffraction peaks could be indexed to the tetragonal crystal system with space group *P4mm*, with lattice parameters values $a$ = 4.0165(6) Å, $c$ = 4.1027(9) Å, and $V$ = 66.184(16) Å$^3$ in good agreement with the reported result.[23] The tick pattern shown at the bottom of the figure is the calculated peak positions for the tetragonal phase. Energy dispersive X-ray analysis on single crystal confirmed the presence of all the cationic elements in agreement with expected stoichiometry within experimental error.

Figure 2 shows the variation of dielectric constant ($\varepsilon_r$) with temperature at different frequencies. The temperature dependence of the dielectric constant exhibited only one dielectric anomaly/peak in the temperature range of investigation. The dielectric peaks observed around $T_m$ ~ 490 K exhibit a clear but weak frequency dispersion. In order to find the order of diffusivity of phase transition, the high temperature side of the dielectric peak was fitted with the modified Curie-Weiss law,

$$1/\varepsilon_r - 1/\varepsilon_{max} = (T - T_m)^\gamma / C, \qquad (1)$$

where $\gamma$ is the diffusivity and $C$ is the Curie-Weiss constant, and $\varepsilon_{max}$ is the maximum value of $\varepsilon_r$ at $T_m$. The value of $\gamma$ is estimated from the slope of log $(1/\varepsilon_r-1/\varepsilon_{max})$ versus log $(T-T_m)$ plots. We obtained the parameter $\gamma$ as 1.56 from fitting the data at 10 kHz, as shown in inset of Fig 2. This confirms the relaxor nature of the system. Similar relaxor nature of the sample has been reported from the inelastic neutron scattering studies.[23]

Figure 3(a) shows the Brillouin spectra in the FSR of 100 GHz at selected temperatures. A Brillouin doublet in the spectra corresponds to the LA mode. In order to precisely determine the Brillouin shift, its full width at half maximum (FWHM) and intensity, we use Voigt functions, where the width of the Gaussian component in the Voigt function is fixed as that of the instrument-function. For the spectrum measured at 363 K, the individual components of the fit are shown in Fig. 3(b). Figure 4 shows the LA mode frequency and its FWHM as a function of temperature. One can see that above 650 K, the frequency and FWHM ($\Gamma_{LA}$) of LA mode exhibit marginal monotonic changes. At a sufficiently high temperature, relaxor ferroelectrics are in the paraelectric phase without any PNRs. In this temperature range the Brillouin shift is expected to show an almost linear temperature dependence without any pronounced change in the FWHM, indicating that the normal anharmonic process governs the lattice dynamical behavior. However, below 650 K, LA mode frequency shows strong softening till $T_{tc}$. One can also see that a substantial increase in its FWHM begins around 525 K and reaches a maximum at 463 K. Note that the Brillouin shift and FWHM show sharp anomalies at $T_{tc}$, representing the occurance of the tetragonal-cubic transition. Earlier X-ray diffraction studies for this composition have suggested the transition temperature to be ~ 490 K.[23] With increasing $x$ content the LA phonon mode has been reported to show a gradual change from broad anomalous feature to a sharp peak across the tetragonal-cubic transition.[8, 9] Therefore, in the present study on $x = 0.15$, the very sharp feature in LA mode across the transition is understandable.

The temperature evolution of the intensity of LA phonon mode, shown in Fig. 5(a), exhibits a substantial change at 650 K and a plateau region below 525 K, followed by an anomaly at $T_{tc} \sim 463$ K. In relaxor ferroelectric materials, the origin of anomalous behavior of acoustic phonon mode has been attributed to PNRs, which begin to form below $T_B$.[8, 20] In the spirit of Landu expression of free energy,[28] the coupling term $F_C$ between the order parameter $Q$ and the strain $\varepsilon$ is usually written in terms of increasing powers of these as,

$$F_C(Q,\varepsilon) = \beta_{ij} Q_i \varepsilon_j + \eta_{ijk} Q_i Q_j \varepsilon_k + \delta_{ijk} Q_i \varepsilon_j \varepsilon_k + \ldots\ldots \quad (2).$$

The first term corresponds to the bilinear coupling between the order parameter $Q$ and $\varepsilon$. The coupling quadratic in order parameter and linear in the strain (electrostrictive coupling) is represented by second term. In relaxor phase, the PNRs are randomly oriented below $T_B$ and the net macroscopic spontaneous polarization $(\langle Q \rangle = 0)$, but $(\langle Q^2 \rangle \neq 0)$. Therefore, one can expect that the only second term would be the dominant factor in the expression of coupling free energy $F_C$, below $T_B$. Further the change in elastic constant owing to the effect of this electrostrictive coupling could be written as[28]

$$\Delta C = C - C^\infty = \eta^2 \langle Q^2 \rangle \chi^{(\varepsilon)} \quad (3)$$

where $\eta$ is the electrostrictive coefficient, $\langle Q^2 \rangle$ is the mean value of squared local polarization, and $\chi^{(\varepsilon)}$ is the clamped susceptibility. In the Landau theory of second-order phase transition $C(T)$ shows a step-like anomaly at the transition temperature because of total cancellation of $\chi^{(\varepsilon)}$ and $\langle Q^2 \rangle$ below the transition temperature. However in case of relaxor, this type of cancellation is not possible since the PNRs nucleate at $T_B$ and local polarization grows gradually with decreasing temperature, which in turn contributes to the decrease of the elastic constant continuously until PNRs saturate. Therefore, the effect of electrostrictive coupling between the order parameter and strain causing a decrement of elastic constant and hence the softening of Brillouin shift with decreasing temperature is understandable. The manifestation of electrostrictive coupling in PZN-PT is reflected in a gradual softening of the LA phonon frequency upon cooling from $T_B$ to $T_{tc}$, shown in Fig. 4. In the case of relaxors (1-$x$)PZN-($x$)PT with lower $x$,[8, 9] PMN-PT,[29] and prototype relaxor PMN[9] the softening of acoustic mode begins at $T_B$. In the present investigation the observation of anomalous behaviors of Brillouin shift, FWHM of LA mode and concurrent change in its intensity clearly suggest the $T_B$ to be ~ 650 K. Further corroborative evidence for $T_B$ has been obtained from the $T$-derivative of the FWHM[30] of the LA mode. Figures 5(b) and 5(c) respectively show the $T$-derivatives of line width $\Gamma$ of the LA phonon and the LA mode frequency $\nu_B$. One can see that the $T$-derivative of FWHM of LA mode (Fig. 5(b)), exhibits a change in slope at ~650 K, confirming it to be the $T_B$. For systems with larger $x$ (7%[8] and 9%[9]), the $T_B$ was found to be above 700 K. The decrease in $T_B$ in the present system could be related to the decrease of degree of relaxer behaviour and enhancement of ferroelectric ordering[9] with increasing Pb content. Therefore, the $T_B$ approaches towards $T_{tc}$. The temperature dependence of FWHM of LA phonon exhibits a λ-type peak at $T_{tc}$ (Fig. 4). This is attributed to the Landau-Khalatnikov-like damping mechanism, which is connected with the bilinear coupling between strain and order parameter $(\langle Q \rangle \neq 0)$.[28] Similar mechanism was found in case other relaxor ferroelectric system PZN-PT with $x = 9\%$[9] and also in $Pb(Sc_{0.5}Nb_{0.5})O_3$,[11] which exhibited relaxor to normal ferroelectric transition. The plateau region of intensity plot of LA phonon will be discussed later.

As mentioned earlier, dynamical relaxations of PNRs give rise to a central peak in the Brillouin spectrum. One can see from Fig. 3 that a central peak, which is named as narrow-CP (nCP), is indeed present at lower temperatures; however, its intensity reduces upon heating and nCP finally disappears above 525 K. As the relaxor ferroelectric materials exhibit evidence of multiple time scales of relaxations,[8,20,30,31] it is worth examining if this system also possesses more than one relaxation times. In order to confirm this, Brillouin measurements were also made using a FSR of 400 GHz. Figure 6(a) shows the Brillouin spectra obtained using 400 GHz FSR at selected temperatures. A broad central peak (bCP) is found at lower temperatures. Upon heating it becomes weak and finally vanishes above $T^*$. Thus the behavior of bCP is similar to that of nCP confirming the existence of two relaxation processes. These two type of CPs are related to relaxation processes associated with polarization flipping in PNRs.[8,31] The slower relaxation has been attributed to strain associated with a non-180° flipping process while the faster relaxation to a 180° flipping process not involving strain. This is due to the reason that the flipping of polarization with strain can be slower than that without strain.[8,31] Further corroborative evidence for existence of more than one relaxation time was found from the fitting of the entire CP (nCP + bCP), which could not be reproduced with single Lorenzian function.[8] Therefore, the appearance of both nCP and bCP indicates the existence of relaxation processes involving multiple time scales. Assuming a Debye relaxation process, both nCP and bCP could be fitted using single Lorenzian functions centered at zero frequency. The fit for the bCP spectrum at 523 K is shown in Fig. 6(b). The temperature dependence of integrated intensity and FWHM of both the CPs is shown in Fig. 7. One can see that as the temperature is reduced from high temperature, the integrated intensities of both the CPs reach maximum around $T_{tc}$. The FWHMs of both the CPs show a decreasing trend toward $T_{tc}$ on approaching from paraelectric to relaxor region. The increase in intensity of CPs and decrease in their line-width with decreasing temperature can be attributed to an increase in PNRs density and enhancement of correlation among them,[9] which in turn is related to a slowing down of the relaxation processes associated with PNRs.

It has been argued that the softening of LA mode below $T_B$ is a consequence of coupling of polarization fluctuation of PNRs with strain fluctuation.[20] In order to obtain the relaxation time of the polarization fluctuation $\tau_{LA}$ by acoustic phonons, the following equation, assuming a single relaxation process,[8, 20] can be used,

$$\tau_{LA}(T) = \Gamma_{tc} / 2\pi(v_\infty^2 - v_B^2) \qquad (4)$$

where $v_\infty$ denotes the Brillouin shift in the high temperature limit, $\Gamma_{tc}$ is the FWHM at the phase transition temperature. We used a linear temperature dependence of $v_\infty$ as shown in Fig. 4, because of the fact that in the high-temperature range ($> T_B$), relaxor ferroelectrics are in the paraelectric phase without any PNRs. In this temperature range, $v$ shows the linear temperature dependence, which is only due to lattice anharmonicity.[20] Below $T_{tc}$ the relaxation time is expected to be strongly influenced by the structural transition. Therefore, $\tau_{LA}$ is calculated for temperatures above 463 K and is shown in Fig. 8. In addition, the relaxation times above $T_{tc}$ can also be determined from both the central peaks (nCP and bCP) assuming Debye-relaxation processes, using the following equation

$$\tau_{CP} = 1/\pi\Gamma_{CP} \qquad (5)$$

Figure 8 also shows the temperature dependence of relaxation time obtained from Eq. (5). One can notice the relaxation time $\tau_{LA}$ obtained from LA phonon mode is of nearly same magnitude as that obtained from bCP. This suggests that the relaxation of PNRs is determined by polarization fluctuations coupled with local strain fluctuations via the LA phonon mode. Furthermore, the calculated relaxation time increases gradually while approaching towards $T_{tc}$ (Fig. 8), indicating a slowing down of the PNRs dynamics in the relaxor phase ($T_{tc} < T < T_B$). Figure 9 shows the temperature dependence of $1/\tau_{bCP}$, clearly exhibiting a linear behaviour below $T_B$, in the vicinity of $T_{tc}$. This implies that the tetragonal-cubic phase transition of PZN-PT is order-disorder type. Such a behaviour is known as critical slowing down of order-disorder transition and the relaxation time given by the following relation:[32]

$$\frac{1}{\tau} = \frac{1}{\tau_o} + \frac{1}{\tau_1}\left(\frac{T-T_{tc}}{T_{tc}}\right) \qquad (6)$$

where, $\tau_1$ is the characteristic time and $\tau_o$ is the relaxation attributed to defects.[33] Using Eq. (6), the best fit for the relaxation time (Fig. 9) yields $\tau_o$=2.3 ps and $\tau_1$=0.28 ps. It is worthy to compare the values of $\tau_1$ obtained for other relaxor system exhibiting similar order-disorder type behaviour in the vicinity of the transition temperature. In case of 70Pb(Sc$_{0.5}$Nb$_{0.5}$)O$_3$-30PbTiO$_3$,[10] the value of $\tau_1$=0.47 ps has been reported. Similarly $\tau_1$= 2.3 ps is found in Pb(Sc$_{0.5}$Ta$_{0.5}$)O$_3$ relaxor.[11] From these results, one can conclude that an order-disorder mechanism contributes to the slowing down of the PNRs dynamics in PZN-PT. It may be mentioned that often relaxation processes have associated activation energy. In order to obtain the activation energy for PNRs relaxation dynamics, the $T$-dependence of bCP relaxation time is analyzed using an Arrhenius plot as shown in inset of Fig. 9. The activation energy turns out to be 236±11 meV. The activation energy for the PNR relaxation dynamics in PZN-PT is comparable to that found in other system such as LiTaO$_3$,[34] and PMN-PT[29] exhibiting a similar order-disorder relaxation dynamics.

Recent acoustic emission studies on PZN and PZN-PT systems have revealed the existence of a characteristic temperature $T^*$ below Burns temperature $T_B$.[26, 27] From these studies, it has been argued that the association of PNRs with local strain fields occurs at $T^*$ but not at $T_B$. Toulouse et al.[35] found the characteristic temperature $T_B$ and $T^*$ of the PZN relaxor system from the analysis of Raman parameters like optical phonon modes and CP. The Burns temperature $T_B$ was argued to be associated with the formation of dynamic PNRs related to the short-lived correlation of off-centered ions, whereas below the intermediate temperature $T^*$ long-lived PNRs appear to have formed, which is related to appearance of permanent correlation of the off-centered ions. Recently, It has been suggested that the $T^*$ could be obtained from the extremum of the temperature derivative of Brillouin shift of LA phonon.[20] Figure 5(c) shows the temperature derivative of Brillouin shift of LA phonon, exhibiting a maximum at 525 K. This suggests that the strain and polarization electrostrictive coupling of PNRs exhibits enhancement at $T^*$. Furthermore, the appearance of plateau region below 525 K in the plot of intensity versus temperature in Fig. 5(a) could be due to a signature of $T^*$ at 525 K, where the statics PNRs begin to form. The other corroborative evidence[8] for the existence of $T^*$ comes from the appearance of CPs below the temperature $T^* \sim 525$ K (Fig. 8). Dkhil *et al* found that most of the complex lead-based relaxor ferroelectrics exhibit nearly the same intermediate characteristic temperature $T^* \sim 500\pm30$ K.[12] Furthermore, from the Brillouin scattering studies on (1-$x$)PZN-$x$PT ($x$ = 4.5 and 9%), $T^*$ was found in the range 500-550 K.[9] In the present studies, $T^*$ turns out to be ~ 525 K.

Figure 10 shows the variation of integrated intensities of bCP in the unpolarized (VV+VH) and depolarized (VH) geometries. In both cases, a sharp increase in intensities around $T_{tc}$ is found. Furthermore, the depolarized intensity is weak suggesting a strongly polarized nature of scattering. The tetragonal phase has a unique polarization axis, hence it is likely that the PNRs in PZN-PT tend to align along that polarization axis while approaching towards tetragonal phase. In fact, this is expected for cubic to tetragonal structural phase transition.[30] Furthermore, the intensity reaches a maximum a little below $T_{tc}$ and then reduces. This suggests that PNRs are still present in the neighbourhood of $T_{tc}$ in the tetragonal phase away from $T_{tc}$, establishing a long range polar order.

## IV. CONCLUSIONS

In conclusion, the temperature dependences of LA phonon and CP in relaxor ferroelectric PZN-PT single crystal have been investigated using Brillouin scattering in the temperature range 300-873 K. Below the Burns temperature $T_B$, softening of mode frequency, followed by an increase in its FWHM is found. At the tetragonal-cubic phase transition temperature ~ 463 K, both LA mode frequency and FWHM exhibit clear anomalies. Slowing down of relaxation time of PNRs is noticed from the narrowing of CPs while approaching the transition from high temperature side. An intermediate characteristic temperature $T^*$ ~525 K is found below $T_B$. The electrostrictive coupling between the strain and polarization fluctuation begins to take place below temperature $T_B$. The relaxation time of CP exhibits critical slowing down upon approaching the tetragonal-cubic transition suggesting the order-disorder nature of phase transition. A comparision of unpolarized and depolarized scattering in cubic phase indicates polarized nature of the broad central peak. This implies that the PNRs tend to align to establish a long range polar ordering.

## ACKNOWLEDGMENTS

We acknowledge Ms. S. Hussain of UGC-DAE CSR, Kalpakkam-node for energy dispersive X-ray analysis of the sample. We also thank Dr. C. S. Sundar for interest in the work and Director IGCAR for encouragement.

**Figure captions**

Figure 1. (Color online) Powder X-rays diffraction pattern of PZN-PT crystal ($x = 0.15$). The tick pattern is the calculated peak positions for the tetragonal phase (*P4mm*). Inset: shows a photograph of a single crystal.

Figure 2. (Color online) Temperature dependence of the dielectric constant ($\varepsilon_r$) of PZN-PT single crystal measured at different frequencies. Inset: $\log(1 / \varepsilon_r - 1 / \varepsilon_{max})$ as a function of $\log(T - T_m)$. Straight line is a linear fit to the data.

Figure 3. (Color online) (a) Brillouin spectra of PZN-PT measured using a FSR of 100 GHz at selected temperatures and (b) Fitted spectrum along with the data at $T = 363$ K. Individual fitted peaks are also shown.

Figure 4. (Color online) Temperature dependences of Brillouin shift and FWHM of the LA phonon mode of PZN-PT single crystal. Solid line is a linear fit to the Brillouin shift data above 750 K, using $\nu_\infty = A - BT$, where $A$ and $B$ are constants. $\Gamma_T$ denotes the FWHM at the phase transition.

Figure 5. (Color online) Temperature dependences of (a) the intensity of LA phonon mode, and (b) Temperature derivative of the LA phonon mode FWHM. (c) Temperature derivative of the LA phonon mode frequency.

Figure 6. (Color online) (a) Brillouin spectra of PZN-PT measured using a FSR of 400 GHz at selected temperatures. (b) Fitted spectrum along with the data at $T = 523$ K. Individual fitted peak is also shown.

Figure 7. (Color online) Temperature dependences of FWHM and the intensity of (a) narrow-CP and (b) broad-CP.

Figure 8. (Color online) Temperature dependences of the relaxation times ($\tau_{LA}$) determined from the LA mode and the relaxation times ($\tau_{ncp}$ and $\tau_{bcp}$) determined from the FWHM of narrow and broad-CP.

Figure 9. (Color online) Plot of $1/\tau_{bCP}$ as a function of temperature $(T-T_{tc})/T_{tc}$; The straight line is a fit to Eq. (6). The inset: shows the logarithmic relaxation time ($\tau$) versus $1/T$. The straightline is a fit to Arrhenius law.

Figure 10. (Color online) Temperature dependences of the integrated intensities of broad central peak measured in unpolarized (VV+VH) and depolarized VH geometries.

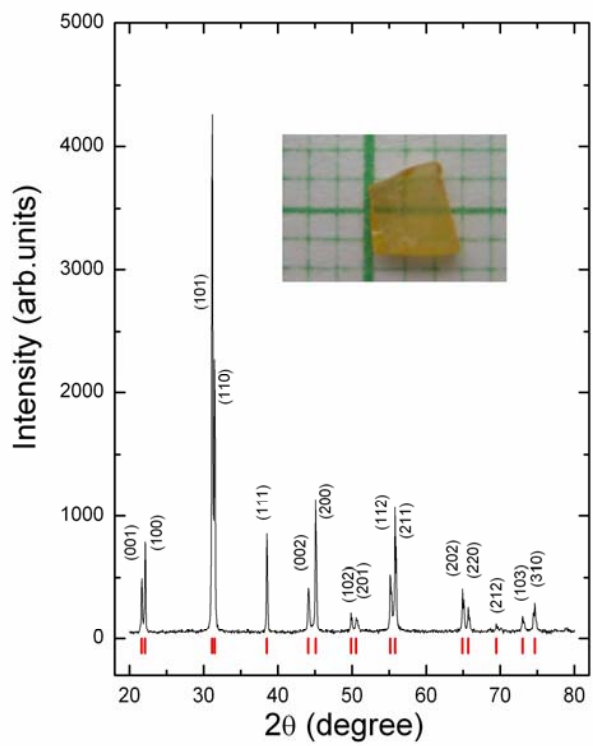

Fig. 1 Mishra *et al*.

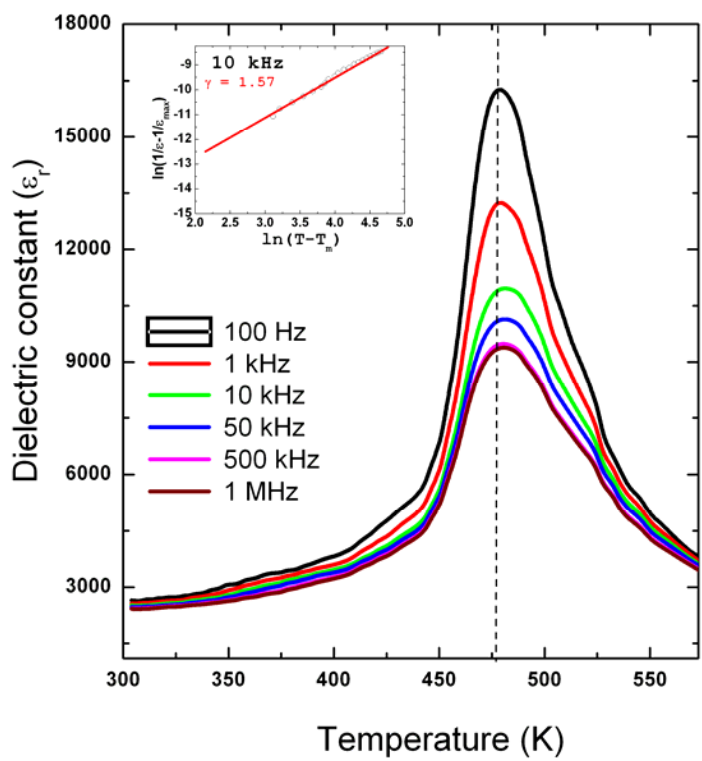

Fig. 2 Mishra *et al.*

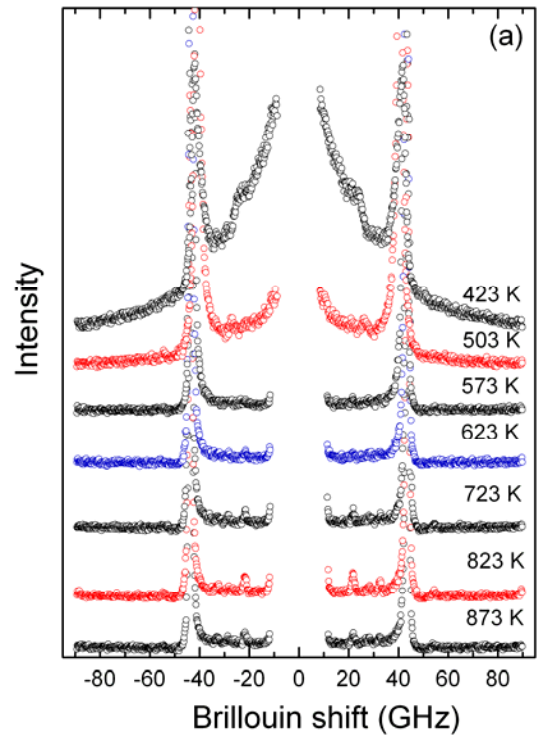

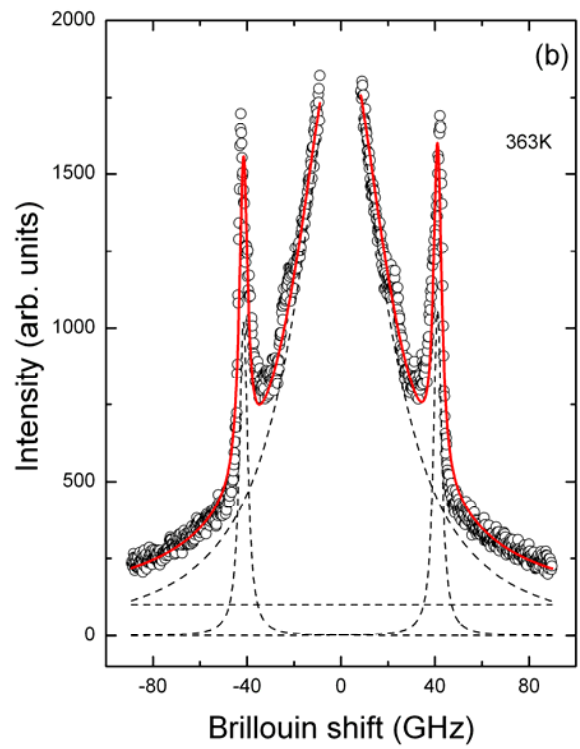

Fig. 3 Mishra *et al.*

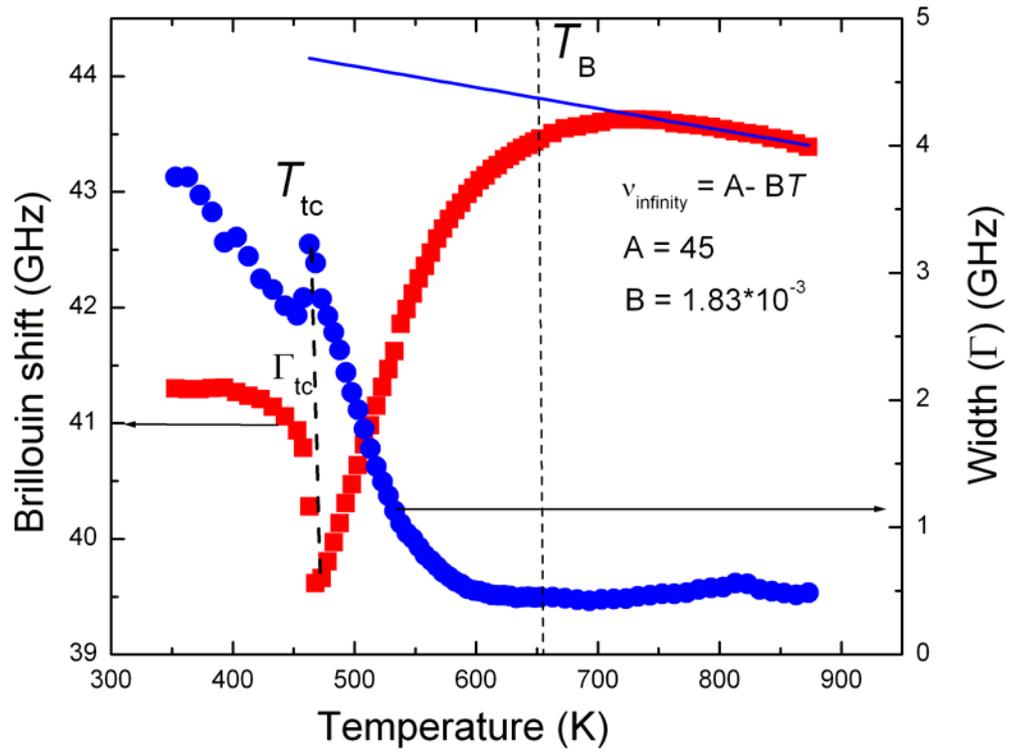

Fig. 4 Mishra *et al.*

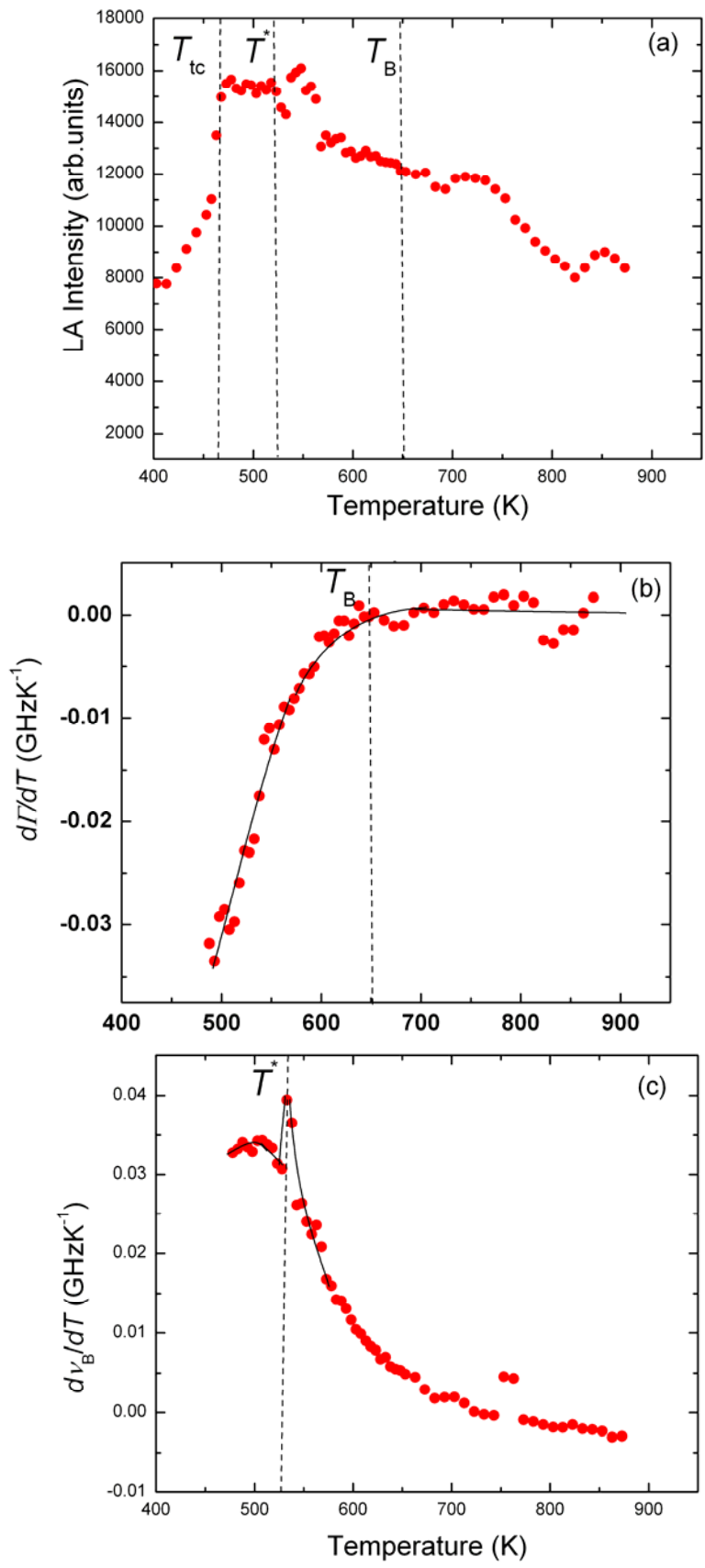

Fig. 5 Mishra *et al.*

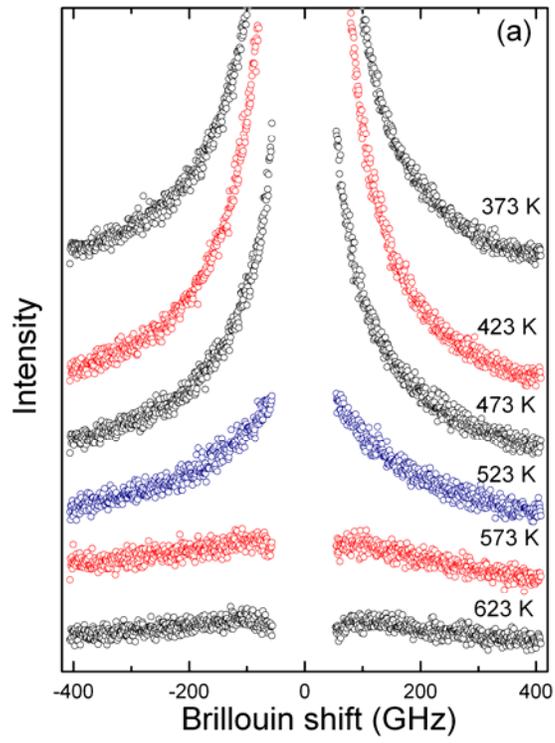

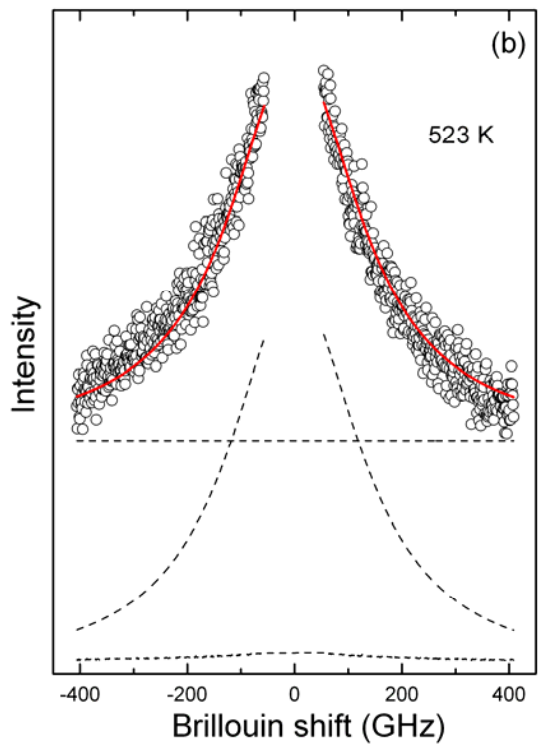

Fig. 6 Mishra *et al.*

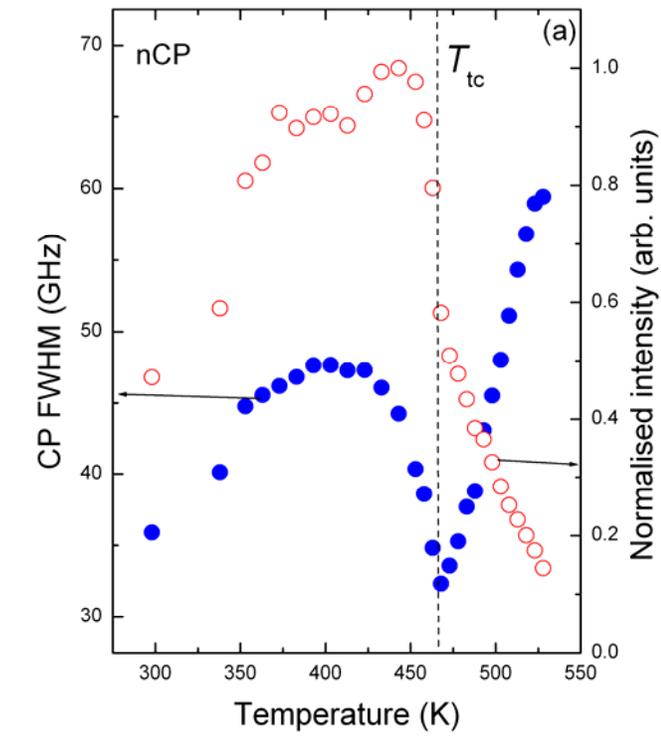

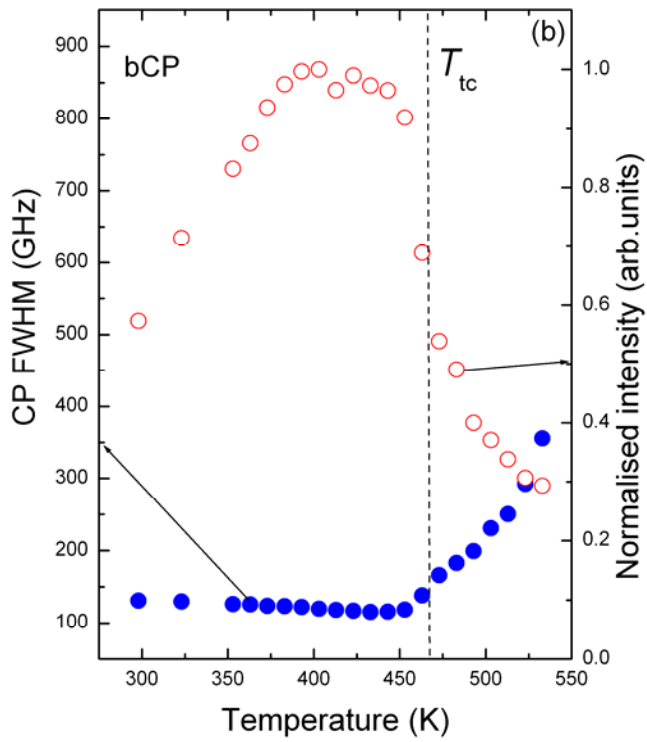

Fig. 7 Mishra *et al.*

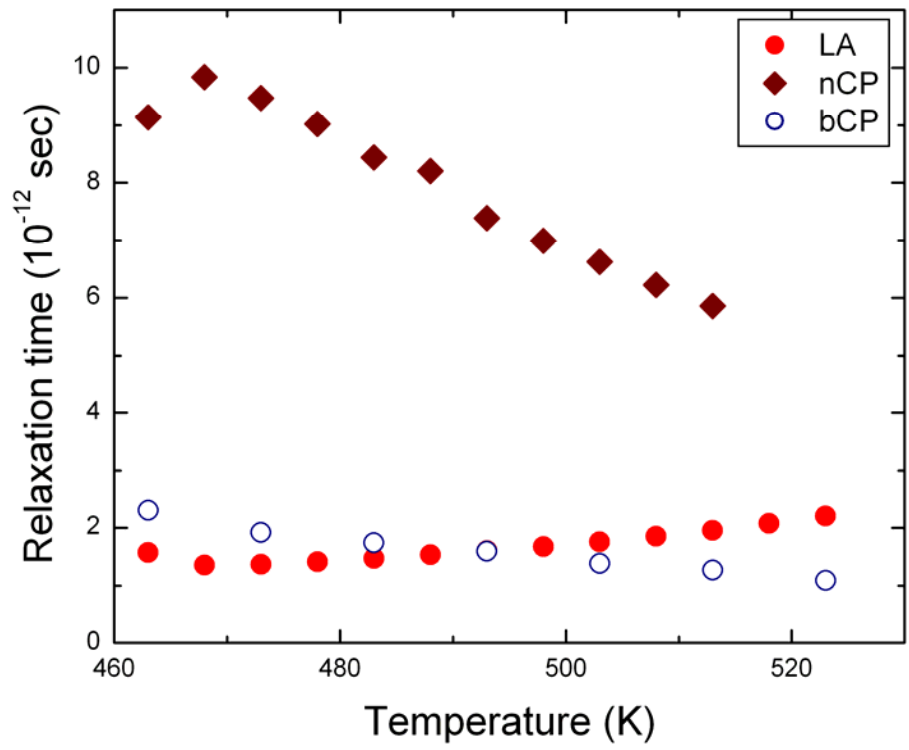

Fig. 8 Mishra *et al.*

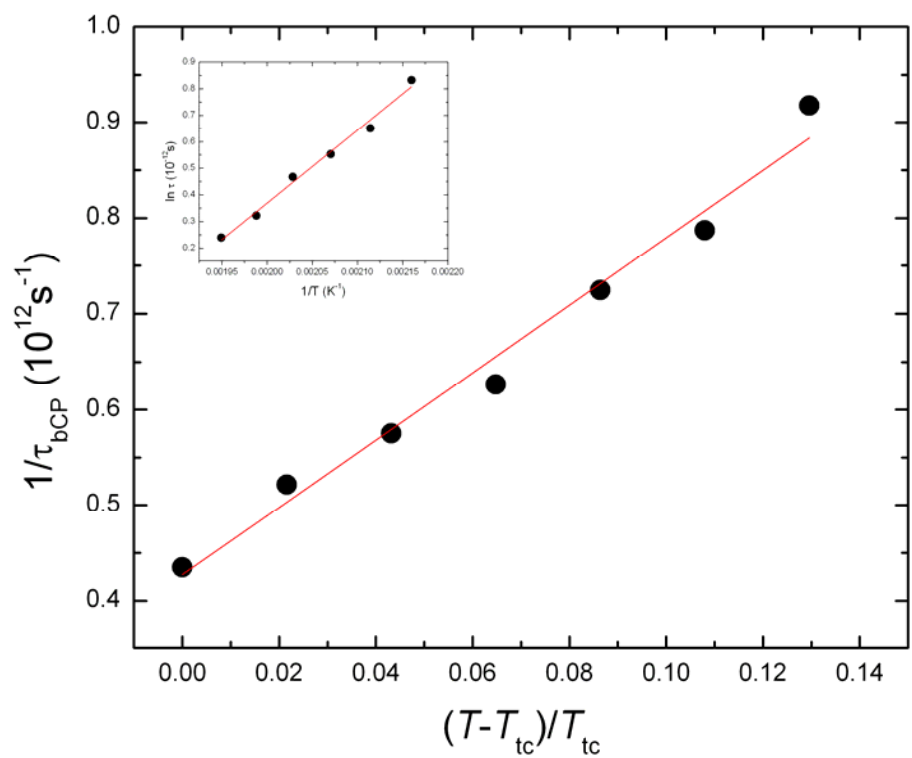

Fig. 9 Mishra *et al.*

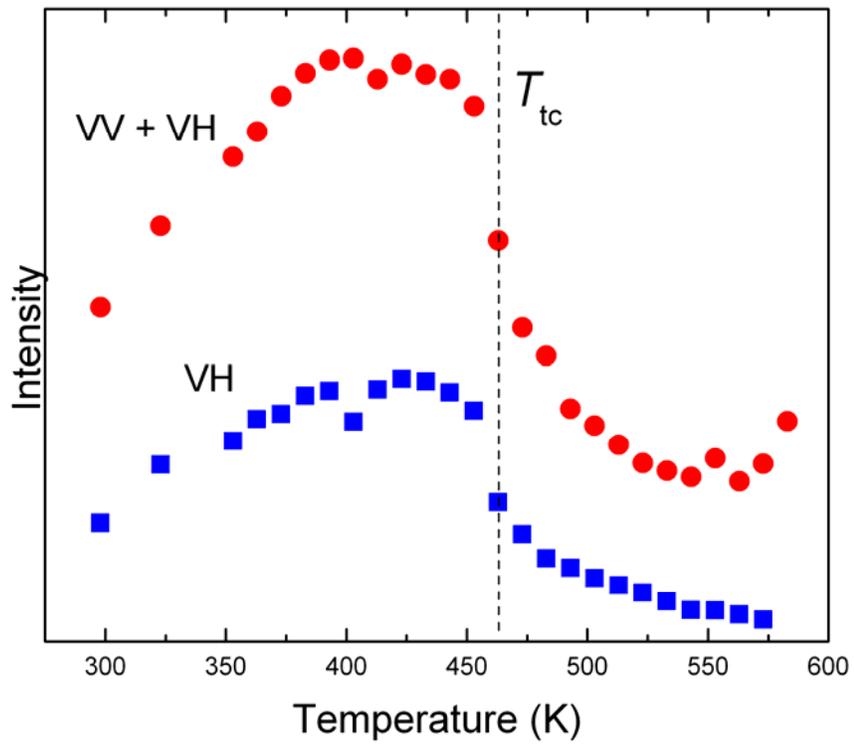

Fig. 10 Mishra *et al.*